\documentclass{SPW9_proc_old}

\usepackage{shortvrb}
\MakeShortVerb{\|}
\def\pdfLaTeX{pdf\kern.06em\LaTeX}
\usepackage{color}	

\newcommand\arcsec{\mbox{$^{\prime\prime}$}}%
\newcommand\farcs{\mbox{$.\!^{\prime\prime}$}}%

\begin{document}

\title{Spectro-Polarimetric Analysis of an Umbral Filament}

\author[1,*]{S.L. Guglielmino}
\author[2]{P. Romano}
\author[3,4]{B. Ruiz Cobo}
\author[1]{F. Zuccarello}
\author[5]{M. Murabito}

\affil[1]{Dipartimento di Fisica e Astronomia ``Ettore Majorana'' -- Sezione Astrofisica, Universit\`a degli Studi di Catania, Catania, I-95123, Italy}
\affil[2]{INAF -- Osservatorio Astrofisico di Catania, Catania, I-95123, Italy}
\affil[3]{IAC -- Instituto de Astrof\'isica de Canarias, La Laguna, Tenerife, E-38200, Spain}
\affil[4]{Departamento de Astrof\'isica, Univ.~de La Laguna, La Laguna, Tenerife, E-38205, Spain}
\affil[5]{INAF -- Osservatorio Astronomico di Roma, Monte Porzio Catone, I-00078, Italy}
\affil[*]{\textit{Email:} salvatore.guglielmino@inaf.it}

\runningtitle{Spectro-Polarimetric Analysis of an Umbral Filament}
\runningauthor{S.~L.~Guglielmino et al.}

\firstpage{1}

\maketitle

\begin{abstract}
High-resolution observations of the solar photosphere have recently revealed the presence of elongated filamentary bright structures inside sunspot umbrae. These features, which have been called umbral filaments (UFs), differ in morphology, evolution, and magnetic configuration from light bridges that are usually observed to intrude in sunspots.

To study an UF observed in the leading sunspot of active region NOAA 12529, we have analyzed spectro-polarimetric observations taken in the photosphere with the spectropolarimeter (SP) aboard the \textit{Hinode} satellite. High-resolution observations in the upper chromosphere and transition region taken with the \textit{IRIS} telescope and observations acquired by \textit{SDO}/HMI and \textit{SDO}/AIA have been used to complement the spectro-polarimetric analysis.

The results obtained from the inversion of the \textit{Hinode}/SP measurements allow us to discard the hypothesis that UFs are a kind of light bridge. In fact, we find no field-free or low-field strength region cospatial to the observed UF. In contrast, we detect in the structure Stokes profiles that indicate the presence of strong horizontal fields, larger than 2500~G. Furthermore, a significant portion of the UF has opposite polarity with respect to the hosting umbra. In the upper atmospheric layers, we observe filaments being cospatial to the UF in the photosphere. We interpret these findings as suggesting that the UF could be the photospheric manifestation of a flux rope hanging above the sunspot, which triggers the formation of penumbral-like filaments within the umbra via magneto-convection.  

\end{abstract}

\section{Introduction}

Light bridges (LBs) are bright features that intrude sunspots from the leading edge of penumbra into the umbra, being usually observed during the decay phase of sunspot evolution \citep[e.g.,][]{Falco:16}. High-resolution observations allow investigating their internal structure, distinguishing between granular or filamentary LBs, and their magnetic configuration. 

Recently, \citet{Kleint:13} have studied unusual curled intruding structures around the preceding sunspot of active region (AR) NOAA~11302, which have been called umbral filaments (UFs). Although at a first glance these features may resemble filamentary LBs, they appear to be different from the latter in morphology and in evolution.  

To assess the properties of these newly discovered features, we have analyzed an UF located inside the umbra of the giant leading sunspot of AR NOAA~12529 \citep{Guglielmino:17}. Using photospheric data acquired by \textit{SDO}/HMI, we have found that this intruding structure has a strong horizontal magnetic field ($\approx 2000$~G), with a portion of the UF characterized by opposite polarity than the hosting sunspot. Considering the cospatial structures seen at chromospheric and coronal heights with \textit{SDO}/AIA, we have proposed that this UF is the counterpart of a flux rope embedded in the solar atmosphere above the umbra.

Here, we describe the analysis of high spectral and spatial spectro-polarimetric observations of the same UF performed by the \textit{Hinode} spectropolarimeter \citep[SP;][]{Lites:13}. The results reinforce the scenario suggested by the analysis of \textit{SDO} data and completely rule out the idea that the UF is a low-field or field-free region, as usual LBs.

\section{Observations and data analysis}

The \textit{Hinode}/SP instrument obtained a single raster scan along the Fe I line pair at 630.2~nm in Fast Mode (pixel size 0\farcs32, step cadence of 3.8~s), covering a field of view (FoV) of about $300 \arcsec \times 162 \arcsec$ that encompassed the whole AR. The scan lasted from 02:21 to 03:24~UT on April~14, 2016, at the time of the passage at the central meridian of the AR ($\mu = 0.96$).

For the analysis, we removed the stray light contamination using a regularization method, based on a principal component decomposition of the Stokes profiles \citep{PCA:13}, using 15 eigenvectors. This led to an enhancement of the continuum contrast in the quiet Sun from 6.9\% in the original data to 12.1\% in the deconvolved data.

We considered a sub-FoV of $130\arcsec \times 130\arcsec$ centered on the preceding sunspot of the AR, as shown in Figure~\ref{fig:UF}. We inverted the spectra dividing this sub-FoV into three regions, identified by different thresholds of the normalized continuum intensity $I_c$ to account for the different physical conditions: quiet Sun ($I_c > 0.9$), penumbra ($0.5 < I_c < 0.9$), and umbra ($I_c < 0.5$). Accordingly, we used the SIR inversion code \citep{SIR:92} with different initial models for the temperature stratification, following \citet{Murabito:16}. Except for the temperature, the physical quantities were assumed to be constant with height. For more details, we defer the reader to \citet{Guglielmino:19}.

\section{Results}

\begin{figure}[t]
\center
\includegraphics[trim= 0 35 0 0, clip,  scale=0.385]{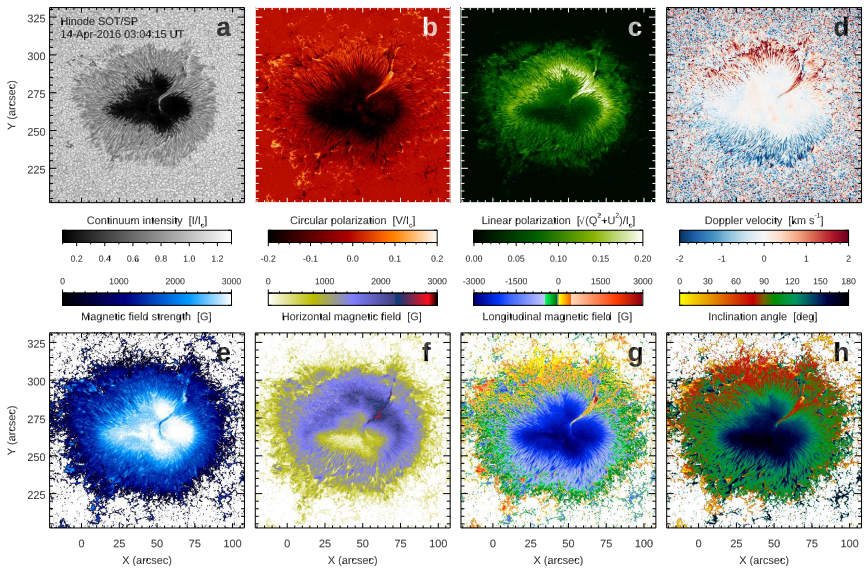} \caption{\textit{Top panels:} Maps of the continuum intensity (\textbf{a}), integrated circular polarization (\textbf{b}) and linear polarization (\textbf{c}), and  Doppler velocity (\textbf{d}) of the preceding sunspot of AR NOAA 12529. \textit{Bottom panels:} Maps of the physical quantities retrieved by the SIR inversion: field strength (\textbf{e}), horizontal (\textbf{f}) and longitudinal (\textbf{g}) component of the vector magnetic field, and inclination angle (\textbf{h}). 
}
\label{fig:UF}
\end{figure}

Figure~\ref{fig:UF} displays the deconvolved maps derived from \textit{Hinode}/SP measurements, as well as the maps of the physical parameters retrieved by the SIR inversion, detailed in the following.

The continuum intensity map (Figure~\ref{fig:UF}\textbf{a}) reveals that the leading edge of the UF corresponds with a penumbral gap, while the trailing edge ends sharply inside the umbra. According to the \textit{Hinode}/SP spatial resolution ($\approx 0\farcs3$ at 630~nm) dataset, there is no granular pattern in the UF. 

The map of integrated circular polarization (Figure~\ref{fig:UF}\textbf{b}) indicates that the signal in a large part of the UF has opposite sign than that of the hosting umbra. Moreover, the integrated linear polarization map (Figure~\ref{fig:UF}\textbf{c}) shows enhanced signal in the UF.

In Figure~\ref{fig:UF}\textbf{d} (Doppler velocity map), we see that  the flow pattern along the UF exhibits a redshift along the part of the structure inside the umbra (normal Evershed flow). Abruptly, a blueshift is observed at the edge of the penumbra, in correspondence of the penumbral gap. 

The maps relevant to the vector magnetic field show that the UF has a magnetic field strength generally larger than 2000~G (Figure~\ref{fig:UF}\textbf{e}), harboring strong horizontal fields. In fact, some patches have horizontal field values larger than 2500~G, stronger than in the surrounding penumbra (Figure~\ref{fig:UF}\textbf{f}). In addition, both the longitudinal field map (Figure~\ref{fig:UF}\textbf{g}) and the inclination angle
map (Figure~\ref{fig:UF}\textbf{h}) confirm the presence of the opposite magnetic polarity, with respect to the sunspot, along the interior of the UF.

It is worth noting that some pixels in the periphery of the UF need a two-component inversion to fit properly the emerging Stokes profiles. In these points, the `umbral' component is weaker than the opposite-polarity `penumbral' component.

\section{Discussion and conclusions}

\textit{Hinode}/SP observations of the UF in AR NOAA~12529 indicate the presence of strong horizontal fields larger than 2500~G, with a significant portion of the UF having opposite polarity than the surroundings. Thus, we can safely confirm that the UF cannot be a kind of granular LB, usually corresponding to field-free or low-field strength regions, or a filamentary LB, usually with a magnetic field of about 1500~G of the same polarity of the hosting sunspot. Rather, our findings hint that the UF is a topologically separated structure.

The characteristics of the UF and its magnetic configuration suggest that it is a penumbral-like feature inside the sunspot umbra. This  penumbral-like nature of UFs might be explained taking into account the presence of a strong enough external horizontal field that is able to activate the penumbral magneto-convective mode \citep{Jurcak:14,Jurcak:17}. In fact, the onset of this physical mechanism has been related with the observations of horizontally inclined magnetic fields in different environments during penumbra formation. This occurs in regular penumbrae \citep{Shimizu:12,Romano:13,Romano:14}, also when they form in regions of ongoing flux emergence \citep{Murabito:17,Murabito:18}, as well as in penumbral-like structures, such as orphan penumbrae, when the emerging flux is trapped by an overlying canopy \citep{Lim:13,Guglielmino:14,Zuccarello:14}.  

This interpretation is further supported by simultaneous images acquired by the \textit{IRIS} satellite that show filaments in the upper atmosphere, highly reminiscent of a flux rope, being cospatial to the UF in the photosphere \citep{Guglielmino:19}. Such a scenario, where the UF is the photospheric counterpart of an overlying flux rope touching the umbra and triggering magneto-convection, bears some resemblance to the \textit{flux sheet} model proposed by \citet{Kleint:13}. New observations from the photosphere to the corona, as well as numerical simulations are required to shed light on the nature of these structures.


\begin{acknowledgements}
Financial support received from the SPW9 LOC to participate to the Workshop in G\"ottingen is gratefully acknowledged. This research was supported by the Istituto Nazionale di Astrofisica (PRIN-INAF 2014), by the Italian MIUR-PRIN grant 2012P2HRCR on The active Sun and its effects on space and Earth climate, by Space Weather Italian COmmunity (SWICO) Research Program, and by the Universit\`a degli Studi di Catania (Piano per la Ricerca Universit\`{a} di Catania 2016-2018 -- Linea di intervento~1 ``Chance''; Linea di intervento~2 ``Ricerca di Ateneo - Piano per la Ricerca 2016/2018''). The research leading to these results has received funding from the European Union's Horizon 2020 research and innovation programme under grant agreements no.~739500 (PRE-EST project) and no.~824135 (SOLARNET project). This work has been partially funded by the Spanish Ministry of Economy and Competitiveness through the Project no.~ESP-2016-77548-C5. 
\end{acknowledgements}


\begin{thebibliography}{}
%
%
%
%
%
%
%
%
%
%
%
%
\bibitem[Cristaldi et al.(2014)]{Cristaldi:14} Cristaldi, A., Guglielmino, S.~L., Zuccarello, F., et al.\ 2014, ApJ, 789, 162 
%
%
%
\bibitem[Falco et al.(2016)]{Falco:16} Falco, M., Borrero, J.~M., Guglielmino, S.~L., et al.\ 2016, Sol.~Phys., 291, 1939 
%
%
%
%
\bibitem[Guglielmino et al.(2019)]{Guglielmino:19} Guglielmino, S.~L., Romano, P., Ruiz~Cobo, B., Zuccarello, F., \& Murabito, M.\ 2019, ApJ, 880, 34
%
%
\bibitem[Guglielmino et al.(2017)]{Guglielmino:17} Guglielmino, S.~L., Romano, P., \& Zuccarello, F.\ 2017, ApJ Lett., 846, L16
%
\bibitem[Guglielmino et al.(2014)]{Guglielmino:14} Guglielmino, S., L., Zuccarello, F., \& Romano, P.\ 2014, ApJ Lett., 786, L22 
%
%
%
\bibitem[Jur{\v c}{\'a}k et al.(2017)]{Jurcak:17} Jur{\v c}{\'a}k, J., Bello Gonz{\'a}lez, N., Schlichenmaier, R., \& Rezaei, R.\ 2017, A\&A, 597, A60
%
\bibitem[Jur{\v c}{\'a}k et al.(2014)]{Jurcak:14} Jur{\v c}{\'a}k, J., Bello Gonz{\'a}lez, N., Schlichenmaier, R., \& Rezaei, R.\ 2014, Publ. Astr. Soc. Jpn, 66, S3 
%
%
\bibitem[Kleint \& Sainz Dalda(2013)]{Kleint:13} Kleint, L., \& Sainz Dalda, A. 2013, ApJ, 770, 74
%
%
%
%
%
%
\bibitem[Lim et al.(2013)]{Lim:13} Lim, E.-K., Yurchyshyn, V., Goode, P., \& Cho, K.-S.\ 2013, ApJ Lett., 769, L18 
%
\bibitem[Lites et al.(2013)]{Lites:13} Lites, B.~W., Akin, D.~L., Card, G., et al.\ 2013, Sol.~Phys., 283, 579 
%
%
%
%
%
%
%
%
\bibitem[Murabito et al.(2018)]{Murabito:18} Murabito, M., Zuccarello, F., Guglielmino, S.~L., \& Romano, P.\ 2018, ApJ, 855, 58 
%
\bibitem[Murabito et al.(2017)]{Murabito:17} Murabito, M., Romano, P., Guglielmino, S.~L., \& Zuccarello, F.\ 2017, ApJ, 834, 76
%
\bibitem[Murabito et al.(2016)]{Murabito:16} Murabito, M., Romano, P., Guglielmino, S.~L., Zuccarello, F., \& Solanki, S.~K.\ 2016, ApJ, 825, 75
%
%
%
%
%
%
%
%
%
\bibitem[Romano et al.(2014)]{Romano:14} Romano, P., Guglielmino, S.~L., Cristaldi, A., et al.\ 2014, ApJ, 784, 10 
%
\bibitem[Romano et al.(2013)]{Romano:13} Romano, P., Frasca, D., Guglielmino, S.~L., et al.\ 2013, ApJ Lett., 771, L3 
%
\bibitem[Ruiz Cobo \& Asensio Ramos(2013)]{PCA:13} Ruiz Cobo, B., \& Asensio Ramos, A.\ 2013, A\&A, 549, L4 
%
\bibitem[Ruiz Cobo \& del Toro Iniesta(1992)]{SIR:92} Ruiz Cobo, B., \& del Toro Iniesta, J.~C.\ 1992, ApJ, 398, 375 
%
%
%
%
%
%
%
%
\bibitem[Shimizu et al.(2012)]{Shimizu:12} Shimizu, T., Ichimoto, K., \& Suematsu, Y.\ 2012, ApJ Lett., 747, L18
%
%
%
%
%
%
%
%
%
%
\bibitem[Zuccarello et al.(2014)]{Zuccarello:14} Zuccarello, F., Guglielmino, S.~L., \& Romano, P.\ 2014, ApJ, 787, 57 
%
\end{thebibliography}

\end{document}